\documentclass[vecphys]{svmult}

\usepackage{makeidx}         
\usepackage{graphicx}        
\usepackage{multicol}        
\usepackage[bottom]{footmisc}

\def\lsim{\raise0.3ex\hbox{$\;<$\kern-0.75em\raise-1.1ex\hbox{$\sim\;$}}} 

\def\gsim{\raise0.3ex\hbox{$\;>$\kern-0.75em\raise-1.1ex\hbox{$\sim\;$}}}

\begin{document}

\title*{Rescuing  $H \to b\bar b$  in  VBF 
at the LHC  by  requiring  a  central  photon
}
\author{Barbara Mele\inst{}}
\institute{INFN, Sezione di Roma,
and Dipartimento di Fisica, Universit\`a La Sapienza, \\
P.le A. Moro 2, I-00185 Rome, Italy\\
\texttt{Barbara.Mele@roma1.infn.it} }
%
%
\maketitle
\begin{abstract}
The LHC potential for a measurement of the Higgs boson coupling 
to the $b$ quark in the standard model is not well established yet. We show that requiring a large transverse momentum photon in the light Higgs boson production via vector-boson fusion
(with subsequent  $H\to b\bar b$ decay) could provide a further handle 
on the $Hb\bar b$ coupling determination, and on the measurement of the 
$HWW$ coupling as well.
\end{abstract}

\section{Introduction}
\label{sec:1}
Once the Higgs boson  will be discovered at the LHC, it will be crucial  to test its properties, and check how well
they fit in the standard model (SM) framework.
Higgs boson couplings to vector bosons, heavy quarks and heavy leptons can in principle be measured by combining informations on different production and decay channels \cite{Reina:2005ae}.

A measurement of the Higgs boson coupling to $b$ quarks 
seems presently quite challenging.
On the one hand, the SM Higgs production channel $b\bar{b}\to H$ 
is overwhelmed by the main production process  $gg\to H$
 at the LHC \cite {maltoni}.
On the other hand,  processes involving the 
$Hb\bar{b}$ coupling via the Higgs decay $H\to b\bar{b}$ 
(for $m_H\lsim 140$ GeV)
seem at the moment hard to manage, due to the large $b$ (and, more generally, jet) background
expected from pure QCD processes.
 The $H\to b\bar{b}$ decay in the
 Higgs production via vector-boson fusion (VBF) has been studied 
 in~\cite{higgsplb}. It gives rise to four-jet final states, out of which
 two jets should be $b$-tagged. Although the VBF final states have  quite distinctive kinematical features (i.e., two forward
jets with a typical  transverse momentum of  order $ M_W$ plus a resonant $b$-jet pair produced centrally),
different sources of QCD backgrounds  and hadronic effects 
presently  make the relevance of this channel 
for a  $Hb\bar{b}$  coupling determination difficult to assess. 
For instance, triggering on  $bbjj$ final states must confront with the corresponding large QCD four-jet trigger rate.
The $Ht\bar{t}$ associated production, where the Higgs boson is 
radiated by a top-quark pair, with subsequent $H\to b\bar{b}$ decay, could also provide a $Hb\bar{b}$ coupling  measurement.
Nevertheless, the
 recent inclusion of more reliable 
QCD background estimate and detector simulation in the corresponding signal analysis~\cite{CMS-TDR}, have
lowered the expectations on the potential of this channel.

Here we report on a further process that could help in 
determining the $H b\bar{b}$ coupling, that was recently studied in 
\cite{nnoi} (where more details can be found).
We consider  the
Higgs boson  production in VBF
in association with a large transverse-momentum  photon 
(i.e., $p_{\rm T}\gsim 20$ GeV) emitted centrally (i.e., with pseudorapidity 
$|\eta_{\gamma}|<2.5$)
\begin{equation}
pp\to H\, \gamma\, jj + X\,\to b \bar b \,\gamma \,jj\, + X\, ,
\label{process}
\end{equation}
where $H$ decays to $b\bar{b}$, and, at the 
parton level, the final QCD partons are
 identified with the corresponding jets $j$. Disregarding the
resonant contribution to the process coming from the $WH\gamma,\,
ZH\gamma$ production,
the  dominant Feynman diagrams  
are the ones involving VBF (as shown in Figure~1, where the Higgs 
decay to $b\bar{b}$ is not shown). Final states 
$b \bar b\, \gamma\, jj\,$  arising from photon radiation 
off one of the two  $b$-quarks arising from the Higgs 
boson decay [via $pp\to H(\to b \bar b\,\gamma )\, jj$]
fall outside the experimental $m_{b\bar b}$  resolution window around 
the $m_H$, due to 
 the requirement of 
a large $p_{\rm T}$ photon. 
%
%
%
\begin{figure}
\centering
\includegraphics[height=7cm]{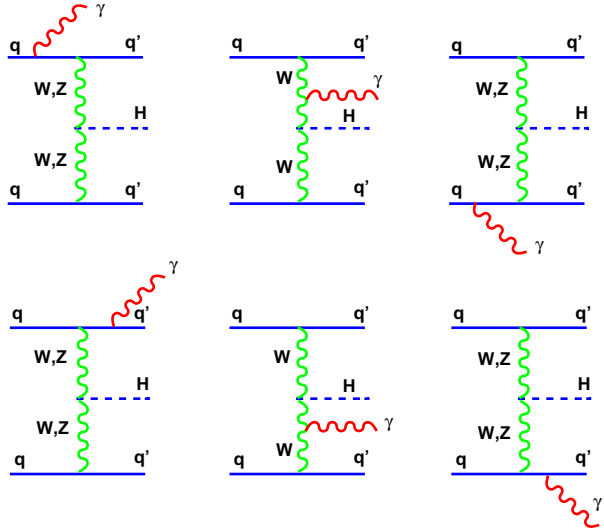}
%
%
\caption{Tree-level $t$-channel Feynman diagrams for  $H$ 
production via $pp\to H\,\gamma\, jj$.}
\label{fig:1}       
\end{figure}
%


\section{Benefits from the central photon}
\label{sec:2}
Adding a central photon to the $pp\to H(\to b\bar b) \; jj$ final state,
despite  a further e.m. fine structure constant $\alpha$
that depletes production rates, gives a number of benefits
\cite{nnoi}
\begin{itemize}
\item
the large (critical) rate for QCD  multi-jet final states that characterizes the background for  $pp\to H(\to b\bar b) \; jj$ is depleted, too, by the electromagnetic coupling when requiring a further photon in the final state;
this is expected to  improve  triggering efficiencies of the detector;
\item
the large  gluonic component entering  the QCD background to the plain 
$b \bar b  \,j j$ final state at parton level does not take part in the
 radiation of  a large $p_{\rm T}$ photon, so making part
 of the potential background to $H \,\gamma \,jj$ {\it inactive};  
 \item
 further dynamical {\it coherence} effects 
dramatically suppress
the radiation of a  photon in the 
irreducible QCD background to  $b \bar b \,\gamma \,jj$, when the photon is central (i.e. emitted outside the typical radiation cone around the initial/final quark legs, for quarks scattered in the $t$-channel) ;
\item
a similar {\it coherence} effect  depletes the
$HZZ$  amplitudes (involving neutral currents) with the respect to the $HWW$  ones (involving charged currents) in Figure~1, increasing
 the relative sensitivity to the $HWW$ coupling 
 in the radiative channel; then, a measurement
of the $b \bar b \,\gamma \,jj$ rate could lead to a combined 
determination of the Higgs boson couplings to  $b$ quarks and $W$
vector bosons, with less contamination from the $HZZ$ coupling 
uncertainties;
\item
the requirement of a central photon 
strongly reduces  the background arising from alternative 
Higgs boson production processes, such as the one
coming from the virtual gluon fusion 
$g^{\ast} g^{\ast} \to H$ diagrams, 
with a  photon radiated off any external 
quark leg.
\end{itemize}
In the following, we will elaborate on a few of the previous items.
\section{Production rates: signal versus background}
\label{sec:3}
 In Table~\ref{tab:1}, the cross section for the signal and irreducible background for the process in Eq.~(\ref{process}) are shown for three values of the Higgs boson mass,
 as independently
obtained by the Monte Carlo event generators ALPGEN~\cite{alpgen},
and MadEvent~\cite{madevent}, with the choice of parameters described in \cite{nnoi}. The following event selection,
that optimizes the significance $S/\sqrt{B}$,
has been applied
 \begin{eqnarray}
&&p_{\rm T}^{j1,b1} \geq 60\, {\rm GeV}, \, \, \, \,\, 
p_{\rm T}^{j2,b2} \geq 30\, {\rm GeV}, \, \,
\, \, \, p_{\rm T}^\gamma \geq 20\, {\rm GeV},  \, \,\,
\Delta R_{ik} \geq 0.7,\, \nonumber \\
&&|\eta_\gamma|\leq 2.5, \, \, \,\,\,
|\eta_b|\leq 2.5, \, \, \,\,\, |\eta_j|\leq 5, \nonumber \\
&&m_{jj} > 800\, {\rm GeV}, \, \, \, \,\,\,\,  
m_H(1-10\%) \leq m_{b \bar b} \leq m_H(1+10\%), \nonumber \\
&&|\Delta \eta_{jj}| > 4, \, \, \, \, \, 
m_{\gamma H} \geq 160\, {\rm GeV},  \, \, \, \, \, \Delta 
R_{\gamma b/\gamma j} \geq 1.2\, , 
 \label{selec}
\end{eqnarray}
where $ik$ is any pair of partons in the final state, and 
$\Delta R_{ik} =\sqrt{\Delta^2\eta_{ik}+\Delta^2\phi_{ik}}$,
with $\eta$ the pseudorapidity and $\phi$ the azimuthal angle.
For comparison, cross sections and irreducible background 
for the plain VBF process are also shown.
\begin{table}
\centering
\caption{Cross sections for  the signal and the irreducible
background for the {\it optimized} event selection, as defined
in Eq.~(\ref{selec}).  The
signal and irreducible background production rates for the plain 
VBF process are also shown, with the same event selection.
}
\label{tab:1}       
\begin{tabular}{llll}
\hline\noalign{\smallskip}
   $m_H $ & 120~GeV  & $\;\; 130$~GeV   &  $\;\;140$~GeV \\
\noalign{\smallskip}\hline\noalign{\smallskip}
$\sigma[H(\to b \bar b)  \gamma jj] \;\;\;\;\;\;\;\;$  
&3.6~fb  &$\;\;\;$2.9~fb &$\;\;\;$2.0~fb  \\
$\sigma[{b \bar b} \gamma jj] $     &$\;$33~fb  
&$\;\;\;\;$38~fb &$\;\;\;\;$40~fb 
 \\ 
 $\sigma[H(\to b \bar b)  jj] $   & 320~fb  & $\;\;\;$255~fb & $\;\;\;$168~fb  
\\ 
$\sigma[{b \bar b} jj] $   & 103~pb  & $\;\;\;$102~pb &$\;\;\;$ 98~pb 
\\
\noalign{\smallskip}\hline
\end{tabular}
\end{table}
In case the usual pattern of QED corrections held,
 the request of a further hard photon would  keep
the  relative weight of signal and background unchanged
with respect to
the $pp\to H\, jj$ case. Indeed, 
the rates for $pp\to H\, \gamma\, jj$ and
its background would be related to a ${\cal O}(\alpha)$ rescaling
of the rates for the $H\, jj$ signal  and its background, respectively,
keeping the $S/B$ ratio approximately stable.
On the other hand, both  the  $H\, \gamma\, jj$ signal and its 
background statistics would 
 decrease according to the  rescaling factor ${\cal O}(\alpha)$.
Consequently, if $(S/\sqrt{B})|_{H(\gamma)\,jj}$
is the signal significance for the VBF process
(with) without a central photon,
  the signal significance for  
$pp\to H\, \gamma\, jj$ would fall down as  
$(S/\sqrt{B})|_{H\gamma \,jj} \sim \sqrt{\alpha}\, 
(S/\sqrt{B})|_{H\,jj}\lsim 1/10\,(S/\sqrt{B})|_{H\,jj} $ 
with respect to the basic VBF process. 
This would question the usefulness of  considering the $H\, \gamma \, jj$
variant of the  $H\, jj$ process, apart from the expected improvement
in the triggering  efficiency of 
the detectors due to the lower  background rates.

In Table~\ref{tab:1}, one can see that
the QED naive expectations  do not necessarily apply when restricted regions of phase space are considered (as discussed in detail in \cite{nnoi}).
We  see that the naive QED  rescaling fails 
for the main background processes $pp\to b \bar b\,( \gamma)\, jj\,$, 
whose rate drops by about a factor 3000 after requiring a central photon,
due to destructive interference ({\it coherence}) effects discussed in \cite{nnoi}.
Since, on the other hand,
the signal cross section roughly follows the naive 
QED rescaling $\sigma_{\gamma}\sim\sigma/100$, the requirement of a central 
 photon gives rise to a dramatic increase  (by more than 
one order of magnitude) in the $S/B$ ratio.
Indeed, in Table~\ref{tab:2},  comparable 
statistical significances for the signal with and without a photon
are obtained, for an integrated  luminosity of $100$~fb$^{-1}$.
The impact of including a few main reducible backgrounds 
for $pp\to b \bar b\, \gamma\, jj\,$ has 
also been studied in \cite{nnoi}, and found to be moderate.
%
\begin{table}
\centering
\caption{Statistical significances 
with the optimized event selection  as defined
in Eq.~(\ref{selec}), for an integrated  luminosity 
of $100$~fb$^{-1}$. The value $\epsilon_b = 60$\% for the 
$b-$tagging efficiency and a Higgs boson event reduction 
by $\epsilon_{b \bar b}\simeq$ 70\%, due to the finite ($\pm$10\%)
$b \bar b$ mass resolution, are assumed. Jet-tagging efficiency and 
photon-identification efficiency are set to 100\%. 
Only the irreducible background is included in $B$.}
\label{tab:2}       
%
%
\begin{tabular}{llll}
\hline\noalign{\smallskip}
   $m_H $ & 120~GeV  & $\;\; 130$~GeV   &  $\;\;140$~GeV \\
\noalign{\smallskip}\hline\noalign{\smallskip}
$S / \sqrt{B}|_{H\gamma\,jj}$ & $\;\;$ 2.6  &$\;\;$ 2.0 &$\;\;$ 1.3  
\\
$S / \sqrt{B}|_{H\,jj}$ &$\;\;$   3.5  &$\;\;$ 2.8 & $\;\;\;$1.9 
\\
\noalign{\smallskip}\hline
\end{tabular}
\end{table}

Apart from enhancing the $S/B$ ratio, coherence effects
in  $pp\to H(\to b\bar b)\gamma\,jj$
  remarkably curb the relative contribution of the 
$ZZ\to H$ boson fusion 
diagrams with respect to the $WW\to H$ ones
(see \cite{nnoi} for further details).
Then,  the 
$H(\to b\bar b)\gamma\,jj$  production at the LHC can 
have a role not only in the determination of  the $Hbb$ coupling, but also for a cleaner determination of  the $HWW$ coupling.

The analysis presented above does not include parton-shower effects.
The latter are expected to  
further differentiate  the signal and background 
final-state topology and composition. A preliminary 
analysis of showering and central-jet veto effects points to 
an improvement of $S / \sqrt{B}$ by about a factor  two \cite{nnoi}. 
The inclusion of  complete showering, hadronization, 
and detector simulations  will be needed to establish the actual 
potential of the process $pp\to H(\to b\bar b)\gamma\,jj$.

\section*{Acknowledgements}
I wish to thank my collaborators Emidio Gabrielli, Fabio Maltoni, Mauro Moretti, Fulvio Piccinini, and 
Roberto Pittau for the enjoyable time I had  in working out with them
the  results discussed above.
This research was partially supported by the 
RTN European Programmes MRTN-CT-2006-035505
(HEPTOOLS, Tools and Precision Calculations for Physics Discoveries at Colliders), 
and  
MRTN-CT-2004-503369 (Quest for Unification).



\end{document}